# Energy-efficient Integrated Sensing and Communication System and DNLFM Waveform

Yihua Ma[1,2], Zhifeng Yuan[1,2], Shuqiang Xia[1,2], Chen Bai[1,2], Zhongbin Wang[1,2], Yuxin Wang[1,2]
[1]State Key Laboratory of Mobile Network and Mobile Multimedia Technology, Shenzhen, China
[2]ZTE Corporation, Shenzhen, China
Email: {yihua.ma, yuan.zhifeng, xia.shuqiang, bai.chen1,wang.zhongbing,wang.yuxin}@zte.com.cn

*Abstract*—Integrated sensing and communication (ISAC) is a key enabler of 6G. Unlike communication radio links, the sensing signal requires to experience round trips from many scatters. Therefore, sensing is more power-sensitive and faces a severer multi-target interference. In this paper, the ISAC system employs dedicated sensing signals, which can be reused as the communication reference signal. This paper proposes to add time-frequency matched windows at both the transmitting and receiving sides, which avoids mismatch loss and increases energy efficiency. Discrete non-linear frequency modulation (DNLFM) is further proposed to achieve both time-domain constant modulus and frequency-domain arbitrary windowing weights. DNLFM uses very few Newton iterations and a simple geometrically-equivalent method to generate, which greatly reduces the complex numerical integral in the conventional method. Moreover, the spatial-domain matched window is proposed to achieve low sidelobes. The simulation results show that the proposed methods gain a higher energy efficiency than conventional methods.

## I. INTRODUCTION

Integrated sensing and communication (ISAC) [1]-[3] is an evolutionary technology in 6G. With ISAC, the future communication system will be able to not only transmit information but also extract valuable information from the physical world. The concept of ISAC is not new, but it becomes popular recently due to the rapid development of wireless communications. The frequency band becomes higher[4], and the antenna number is also greatly increased. Moreover, the computing power is improved as the baseband supports higher and higher throughput, and edge computing is expected to be implemented.

In terms of the integration level, ISAC is mainly divided into 3 levels [2][3]: coexistence, cooperation, and joint design. Coexistence is the lowest level of integration as two sub-systems transmit wireless signals in the same resources. Interference management is required to ensure performance. A higher level is cooperation, which means two sub-systems dynamically sense the spectrum and find suitable resources to transmit. The highest level is the joint design. The two sub-systems are designed as one. In this way, the spectrum, radio-frequency (RF) chain, and processing unit are expected to be shared to achieve the integration gain.

In the joint design of ISAC, the orthogonal frequency division multiplexing (OFDM) based communication hardware can be used to send the sensing signals. In radar, linear frequency modulation (LFM) signal is widely used. In ISAC, LFM can be generated via digital methods. The LFM signal can also be flexibly embedded in the OFDM symbols [5]. Apart from the dedicated sensing waveform, reusing the existing OFDM signal is also an option. The spatial domain can be utilized, and the waveform over multiple antennas is optimized [6]. Also, the unused sub-carriers can be filled with optimized data [7] to ensure the sensing performance. When the sensing task has a high priority, dedicated sensing resources should be preferred. Otherwise, the optimized joint design can be a better option.

This paper considers using sensing reference signal, which is similar to the existing position reference signal (PRS). Just like PRS, sensing signals can also be reused as communication reference signals. However, the existing reference signal designs did not consider sidelobe suppression. As a comparison, in radars, NLFM has been proposed to avoid mismatch loss of windows [8], which is energy-efficient as the SNR degradation resulted from the mismatch is avoided. Moreover, it can be used to generate arbitrary spectrum shape while maintaining the constant modulus feature, which also increases the amplifier transmission efficiency. However, NLFM requires an accurate numerical integral of an inverse function that cannot be expressed. It is not suitable for ISAC as ISAC needs to flexibly adjust the bandwidth, sub-carrier spacing, and time duration according to the sensing service requirement. OFDM-NLFM [9] was also designed for radar, which is low-complexity but very inaccurate as it simply uses multiple short LFM signals to imitate one NLFM signal.

To achieve high energy efficiency, this paper proposes to use the sensing signal with matched transmitting and receiving windows in the ISAC system. The matched filter processing maximizes the receiving signal power to noise ratio (SNR) and increases energy efficiency. The idea is applied in both the time and frequency domain to obtain a more clear Doppler and range detection results. To efficiently gain constant modulus, a novel discrete NLFM (DNLFM) is proposed to better fit the OFDM-based system. DNLFM achieves an efficient generation via few Newton iterations and a simple geometrically-equivalent method, which is much faster than conventional NLFM [8] and much more accurate than OFDM-NLFM [9]. Moreover, the matched window idea is first introduced into the spatial domain, which shows a great reduction in peak sidelobe ratio (PSLR) and integrated sidelobe ratio (ISLR). The simulation results show that the proposed method gains a nearly 3 dB SNR gain in the time-frequency domain and more than 20 dB PSLR and ISLR gains.

This paper is organized as follows. Section II introduces the transmitting and receiving model. Section III proposes the matched window design in time, frequency, and spatial

This work is supported by the National Key Research and Development Program of China (No.2021YFB2900200).

domain. In Section IV, DNLFM is proposed with detailed derivations. Section V shows the simulation results which verify the proposed methods. Section IV briefly concludes this work. In this paper, $(\cdot)^T$, and $(\cdot)^H$ denote transpose, and Hermitian transpose of a matrix or vector. $\otimes$ represents the Kronecker product.

## II. SYSTEM MODEL

### A. Transmitting Model

Assume that the ISAC system sends a frequency-domain sensing OFDM symbol $\mathbf{x}_f \in \mathbb{C}^{N \times 1}$, where $N$ is the number of effective sub-carriers. The sub-carrier spacing is $\Delta f$, and the bandwidth is thus $B = N\Delta f$. It is converted into the time domain and transmitted. The time domain signal is $\mathbf{x}_t = \mathbf{F}_N \mathbf{x}_f$, where $\mathbf{F}_N$ is the $N$-point discrete Fourier transform (DFT) matrix. $M$ copies of $\mathbf{x}_t$ with a CP of $\mathbf{x}_{CP}$ are sent as

$$\mathbf{x} = [\mathbf{x}_{CP}^T, \mathbf{x}_t^T, \mathbf{o}^T, \mathbf{x}_{CP}^T, \mathbf{x}_t^T, \mathbf{o}^T, ..., \mathbf{x}_{CP}^T, \mathbf{x}_t^T, \mathbf{o}^T]^T, \quad (1)$$

where $\mathbf{o}$ is a zero vector, and these zero occasions can be used to transmit communication signals.

The ISAC system has $L$ transmitting antennas and $L$ receiving antennas. Both antennas form a uniform linear array with a spacing of half wavelength. The ISAC scans the space at angle $\theta_{l'}$, where $l' = 0, 1, ..., L'-1$, and $L'$ can be larger than $L$. The multi-antenna transmitting signal is

$$\mathbf{X} = \mathbf{a}(\theta_{l'})\mathbf{x}^T,$$
$$\mathbf{a}(\theta_{l'}) = [e^{j\pi\cos\theta_{l'} \cdot 0} \quad e^{j\pi\cos\theta_{l'} \cdot 1} \quad ... \quad e^{j\pi\cos\theta_{l'} \cdot (L-1)}]^T. \quad (2)$$

### B. Receiving Model

After removing CPs, extracting OFDM symbols and $N$-point DFT, the received signal becomes $\mathbf{Y} \in \mathbb{C}^{N \times M \times L}$, whose element $Y_{n,m,l}$ represents the frequency-domain data at the $n$-th sub-carrier of the $m$-th OFDM symbol in the $l$-th antenna, where $l = 0, 1, ..., L-1$, $m = 0, 1, ..., M-1$, and $n = 1, 2, ..., N-1$. With the frequency-domain data, the channel information $\mathbf{H} \in \mathbb{C}^{N \times M \times L}$ can be obtained via

$$\mathbf{H} = [H_{n,m,l}], H_{n,m,l} = Y_{n,m,l}/(x_f)_n. \quad (3)$$

The 3D matrix $\mathbf{H}$ carries the sensing information of range, Doppler, and angle. The sensing information can be extracted via DFT or inverse DFT in each dimension. Since the transmitting signal is directional, the receiver only requires one spatial-domain DFT-like vector to implement matched filtering. At the angle $\theta_{l'}$, a 2D detection matrix $\mathbf{A}(\theta_{l'}) \in \mathbb{C}^{N \times M}$ can be obtained as

$$\mathbf{A}(\theta_{l'}) = [A'_{n,m}(\theta_{l'})],$$
$$A_{n,m}(\theta_{l'}) = \frac{1}{\sqrt{MNL}} \sum_{l=0}^{L-1}\sum_{j=0}^{M-1}\sum_{i=0}^{N-1} H_{i,j,l} e^{j2\pi\frac{in}{N}} e^{-j2\pi\frac{jm}{M}} e^{j\pi\cos\theta_{l'}l}. \quad (4)$$

After scanning each angle $\theta_{l'}$, a 3D detection result can be obtained as

$$\mathbf{B} = [B_{n,m,l'}], B_{n,m,l'} = A_{n,m}(\theta_{l'}). \quad (5)$$

The target is detected by correctly finding the main lobe of range, Doppler, and angle in $\mathbf{B}$.

## III. ENERGY-EFFICIENT ISAC WITH MATCHED WINDOWS

### A. Time-Frequency Matched Window

When $\mathbf{x}_f$ is a frequency-domain constant-modulus vector, Equation (3) can also be written as

$$\mathbf{H} = [H_{n,m,l}], H_{n,m,l} = Y_{n,m,l}(x_f)_n^H / ((x_f)_n(x_f)_n^H). \quad (6)$$

Combined with Equation (4), the detection processing is a matched filtering of different delays and Doppler frequencies. The correlation peaks of matched filtering indicate the estimated range and Doppler of the target. Although matched filtering maximizes the signal-to-noise ratio (SNR), the sinc function sidelobes of LFM lead to false detection especially when there are near-far effects among sensing targets. Therefore, in practice, a receiving window is necessary. With a window function $\mathbf{w}(a) \in \mathbb{C}^{a \times 1}$, Equation (4) becomes

$$\mathbf{A}'(\theta_{l'}) = [A'_{n,m}(\theta_{l'})], A'_{n,m}(\theta_{l'}) = \frac{1}{\sqrt{MNL}}$$
$$\sum_{l=0}^{L-1}\sum_{j=0}^{M-1}\sum_{i=0}^{N-1} H_{i,j,l} w_i(N) e^{j2\pi\frac{in}{N}} w_j(M) e^{-j2\pi\frac{jm}{M}} w_l(L) e^{j\pi\cos\theta_{l'}l}. \quad (7)$$

However, with only the receiving window, the receiver processing is no longer a matched filter, which leads to mismatch loss. This loss can be calculated by

$$SNR_{loss}(dB) = 10\log_{10}\left(\frac{L_W^2}{L_W}\right) - 10\log_{10}\left(\frac{(\sum w(L_W))^2}{\sum w^2(L_W)}\right)$$
$$= 10\log_{10}\left(\frac{L_W \sum w^2(L_W)}{(\sum w(L_W))^2}\right), \quad (8)$$

where $L_W$ is the window length in the time or frequency domain. Using Cauchy-Schwarz inequality, this SNR loss is always non-negative.

To avoid the mismatch loss, part of the window should be added at the transmitter side as

$$\mathbf{x}_w = [\sqrt{w}_0(M)[\mathbf{x}_{CP,w}^T, \mathbf{x}_{t,w}^T], \mathbf{o},..., \sqrt{w}_{M-1}(M)[\mathbf{x}_{CP,w}^T, \mathbf{x}_{t,w}^T], \mathbf{o}]^T,$$
$$\mathbf{x}_{t,w}^T = \mathbf{F}_N(\sqrt{\mathbf{w}}(N) \otimes \mathbf{x}_f), \quad (9)$$

where $\mathbf{x}_{CP,w}$ is the CP of $\mathbf{x}_{t,w}$.

The receiver processing now becomes

$$\mathbf{A}'(\theta_{l'}) = [A'_{n,m}(\theta_{l'})], A'_{n,m}(\theta_{l'}) = \frac{1}{\sqrt{MNL}}$$
$$\sum_{l=0}^{L-1}\sum_{j=0}^{M-1}\sum_{i=0}^{N-1} H_{i,j,l} \sqrt{w}_i(N) e^{j2\pi\frac{in}{N}} \sqrt{w}_j(M) e^{-j2\pi\frac{jm}{M}} w_l(L) e^{-j\pi\cos\theta_{l'}l}.$$
$$(10)$$

Equations (9) and (10) use matched windows and form a matched filter. When the transmit power is the same for $\mathbf{x}_t$ and $\mathbf{x}_{t,w}$, the matched window gains a higher receiving SNR. Assume that LFM is used as the sensing reference signal. When the window is added to the frequency domain data, the modulus of the time-domain signal is no longer constant. The consequence is that the power amplifier (PA) has a larger de-rating power to prevent non-linear distortion, which affects the transmitting efficiency. This paper further proposes DNLFM to solve this problem in the next section.

## B. Spatial-domain Matched Window

The effect of spatial-domain matched window differs from that of time and frequency. The transmitting and receiving angle spectrum can be modeled as

$$A(\theta) = \frac{1}{L}\sum_{j=0}^{L-1}\left(\left(\sum_{i=0}^{L-1}e^{-j\pi\cos\theta_t i}e^{j\pi\cos\theta j}\right)e^{j\pi\cos\theta j}w_j(L)e^{-j\pi\cos\theta_r j}\right)$$

$$= H_0\left(\frac{1}{\sqrt{L}}\sum_{i=0}^{L-1}e^{j\pi(\cos\theta-\cos\theta_r)l}\right)\left(\frac{1}{\sqrt{L}}\sum_{j=0}^{L-1}w_j(L)e^{j\pi(\cos\theta-\cos\theta_r)l}\right)$$

$$= H_0 A_T(\theta)A_R(\theta), \quad (11)$$

where $H_0$ is the wireless channel from the first transmit antenna to the first receive antenna.

It can be seen that the final angle spectrum is a dot product of the transmitting and receiving beam pattern. When the matched window is used, Equation (11) becomes

$$A'(\theta) = \frac{1}{L}\sum_{j=0}^{L-1}\left(\left(\sum_{i=0}^{L-1}e^{-j\pi\cos\theta_t i}e^{j\pi\cos\theta j}\right)e^{j\pi\cos\theta j}w_j(L)e^{-j\pi\cos\theta_r j}\right)$$

$$= H_0\left(\frac{1}{\sqrt{L}}\sum_{i=0}^{L-1}\sqrt{w_i(L)}e^{j\pi(\cos\theta-\cos\theta_r)l}\right).$$

$$\left(\frac{1}{\sqrt{L}}\sum_{j=0}^{L-1}\sqrt{w_j(L)}e^{j\pi(\cos\theta-\cos\theta_r)l}\right)$$

$$= H_0 A'_T(\theta)A'_R(\theta). \quad (12)$$

Unlike the time-frequency matched window, the spatial-domain matched window is not to avoid the mismatch loss. As shown in Equation (12), both transmitting and receiving beamforming are not matched filters of the spatial channel.

The gain of using the matched window of Equation (12) over the receiving window of Equation (11) is the matching of angle spectrum. When the transmitting and receiving beam patterns are not matched, the receiving mainlobe will multiply the transmitting first sidelobe, which leads to large first sidelobes in the final detection result. As a comparison, the matched angle spectrum suppresses the same sidelobes twice, which therefore achieves a better PSLR and ISLR.

## IV. DNLFM WAVEFORM DESIGN

### A. DNLFM Waveform

To achieve the time-domain constant modulus and the frequency-domain window weighted modulus, the use of NLFM is first introduced into ISAC. NLFM was designed for radar [8], but it has not been widely as LFM gains a much lower complexity and cost for hardware design. Using LFM in radar, the generation can be realized via a simple voltage-control-oscillator, and the baseband only requires a small bandwidth due to the down-sampling of the signal after mixing with the local LFM. However, with high-speed baseband processing provided by the communication nodes, NLFM become competitive for ISAC.

In radar, conventional NLFM was generated via integrating an inverse function [8]. However, this inverse function cannot be expressed in most cases, which makes the generation can only be done via numerical calculations. The accuracy is decided by numerical granularity, which requires a very high complexity. It also becomes difficult for the ISAC system to flexibly switch different window shapes and OFDM parameters. In existing works of OFDM-based NLFM [9], the NLFM signal is generated via multiple sub-chirps. Each sub-chirp is an LFM signal occupying only partial bandwidth. OFDM-NLFM is obtained via approximating curves by polylines. Unlike existing OFDM-NLFM works, this paper proposes a novel discrete NLFM (DNLFM) waveform, which gains a more accurate window shape. Also, an efficient formula-based generation is proposed for digital baseband to implement.

### B. Formula-based DNLFM Generation

In the first step, a desired window weighting function is selected as $V^2(f)$, where $-B/2 < f < B/2$. The corresponding time domain signal is constant-modulus with a phase of $\phi(t)$. According to [8], the stationary-phase principle suggests

$$\Phi''(f) \approx k \cdot V^2(f). \quad (13)$$

Integrating Equation (13) into

$$\Phi'(f) = \int_{-B/2}^{f}\Phi''(x)dx. \quad (14)$$

The group time delay function [8] is then obtained as

$$T(f) = -\frac{1}{2\pi}\Phi'(f). \quad (15)$$

When $f = B/2$, $T(f)$ should be $T_{sym}$, which is the OFDM symbol time. Therefore, $k$ is obtained as

$$k = -2\pi T_{sym}\Big/\int_{-B/2}^{B/2}V^2(f)df. \quad (16)$$

The conventional NLFM [8] generation method uses the inverse function of $T(f)$ and obtains the phase function via

$$\phi(t) = 2\pi\int_0^t f(x)dx, f(t) = T^{-1}(f). \quad (17)$$

The inverse function is usually not available in the form of a math formula even when a simple Hann window is used. As a comparison, OFDM-NLFM [9] directly approximates $f$ via multi-segment lines, and each line can be expressed via an LFM formula. OFDM-NLFM simplifies the generation but inevitably introduces approximation errors.

To enable fast and flexible generation in the ISAC system, this paper proposes a formula-based way, which does not require any function inversion and numerical integral calculation. In the digital ISAC system, the OFDM sampling points $t_n$ are considered, which means

$$T(f_n) = -\frac{1}{2\pi}\Phi'(f) = t_n, t_n = nT/N, n = 0,1,...,N-1. \quad (18)$$

The corresponding frequency $f_n$ can be calculated via the Newton iterations of

$$f_{n,i+1} = f_{n,i} - \frac{\Phi'(f_{n,i-1}) + 2\pi t_n}{\Phi''(f_{n,i-1})}, f_{n,0} = 0. \quad (19)$$

The discrete samples of the integral in Equation (17) can be calculated by a geometrical method using the same area, which only requires to integrate $T(f)$ instead of $T^{-1}(f)$. According to the symmetry of the window coefficients, $f_n = 0$ when $t_n = T/2$. When $t_n \leq T/2$ and $f_n \leq 0$, it equals the minus of the sum of a rectangle area and an integral of $T(f)$ as

$$\phi(t_n) = -2\pi\left(\left(-f_n t_n\right) + \int_{-\frac{B}{2}}^{f_n}T(x)dx\right)$$

$$= 2\pi\left(f_n t_n - \int_{-\frac{B}{2}}^{f_n}T(x)dx\right), t_n \leq \frac{T}{2}. \quad (20)$$

When $t_n > 0$ and $f_n > 0$, the positive-frequency part can also be calculated via a rectangle area minus an integral of $T(f)$ as

$$\phi(t_n) - \phi\left(\frac{T}{2}\right) = 2\pi\left(f_n\left(t_n - \frac{T}{2}\right) - \int_0^{f_n} T(x)dx\right), t_n > \frac{T}{2}. \quad (21)$$

As $\phi(T/2)$ is obtained by Equation (20), Equation (21) becomes

$$\phi(t_n) = 2\pi\left(f_n\left(t_n - \frac{T}{2}\right) - \int_0^{f_n} T(x)dx + f_n\frac{T}{2} - \int_{-\frac{B}{2}}^{0} T(x)dx\right)$$

$$= 2\pi\left(f_n t_n - \int_{-\frac{B}{2}}^{f_n} T(x)dx\right), t_n > 0. \quad (22)$$

Till this step, Equations (20) and (22) can be combined into one for any $t_n$ as

$$\phi(t_n) = 2\pi f_n t_n - 2\pi\int_{-\frac{B}{2}}^{f_n} T(x)dx$$

$$= 2\pi f_n t_n + \Phi(f_n) - \Phi(-B/2). \quad (23)$$

The OFDM-based constant-modulus waveform can then be decided by $\exp(j\phi(t_n))$, $n = 0, 1, 2, ..., N-1$. The proposed formula-based DNLFM requires twice integrals of the target spectrum window $V^2(f)$, which are available for many window functions like Hann window, Hamming window, and Blackman Window. As the twice integrals can be obtained and stored in the hardware, only Equations (19) and (23) are required to generate DNLFM, which is simple. The convergence of Equation (19) is also very fast, which is also verified by simulations.

## V. NUMERICAL RESULTS.

In simulations, an OFDM-based ISAC system is assumed, and the related parameters are listed in Table I. The cell average constant false alarm rate (CA-CFAR) method [10] is used to detect the target, and the detection algorithm settings are also shown in Table I. DNLFM can be used to generated arbitrary spectrum shape, and in the simulation the commonly used Hamming window function is assumed if there is no additional explanation.

TABLE I. SIMULATION PARAMETERS

| Name | Value |
| --- | --- |
| Carrier frequency | 12 GHz |
| Sub-carrier spacing | 60 kHz |
| Effective sub-carrier number, $N$ | 1024 |
| Sensing symbol number, $M$ | 128 |
| Antenna number, $K$ | 32 |
| Simulation oversampling rate | 4 |
| Iteration number in DNLFM generation | 10 |
| Segment number of OFDM-NLFM | 16 |
| Alarm rate of CA-CFAR | $10^{-3}$ |
| Guard Cell Width of CA-CFAR | 8 |
| Training Cell Width of CA-CFAR | 16 |

### A. Time-frequency SNR Gain

To show the time-frequency SNR gain brought by the matched window, the successful detection probability against SNR is simulated. As shown in Fig. 1(a), both the frequency-domain matched-window LFM and the DNLFM achieve around 1.40 dB and 1.45 dB SNR gain over that of conventional LFM at a successful probability of 0.9. These channel gains are close to the theoretic value calculated by Equation (8), 1.35 dB. Moreover, when the time-domain matched window is further added, both enhanced schemes gain around 2.85 dB gain, which shows a great energy efficiency gain of the proposed DNLFM.

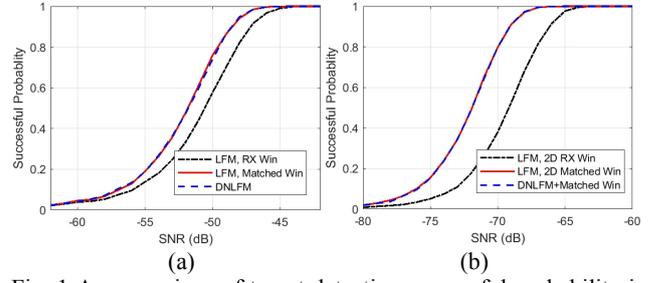

Fig. 1 A comparison of target detection successful probability in the (a) range-domain and (b) range-Doppler domain.

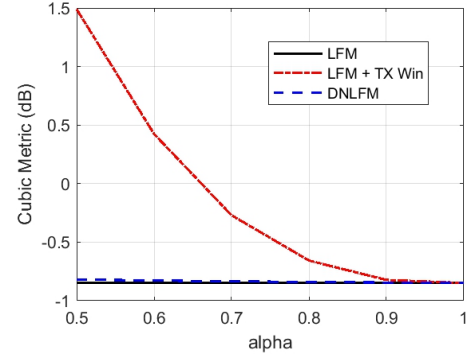

Fig. 2 The cubic metrics of different sensing waveforms.

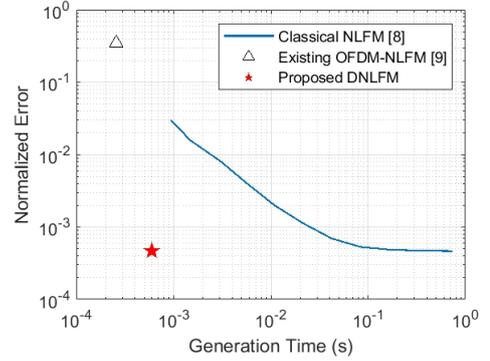

Fig. 3 The generation efficiency comparison between NLFM, OFDM-NLFM and the proposed DNLFM.

Fig. 2 shows the cubic metric of different waveforms. The cubic metric validly predicts the power de-rating of PA according to the 3GPP proposal [11], which is calculated by

$$CM(dB) = \frac{20\log_{10}(rms(v_{norm}^3(t))) - 1.52}{1.56}, \quad (24)$$

where $v_{norm}(t)$ is the normalized time-domain signal. In Fig. 2, $\alpha$ is introduced to control the shape of a window with

$$W_n(N) = \alpha - (1-\alpha)\cos\left(2\pi\frac{n}{N}\right). \quad (25)$$

As the window coefficients are no less than 0, $0.5 \leq \alpha \leq 1$. When $\alpha = 0.5$, it is a Hann window. When $\alpha = 0.54$, it is Hamming window. When $\alpha$ approaches 1, the window shape becomes a rectangle window. Compared with directly adding a window to LFM, DNLFM avoids 2.3 dB and 1.9 dB PA re-rating loss when Hann and Hamming windows are used. The PA de-rating can gain is larger when the window shape is farther away from a rectangle window. This result shows that the proposed DNLFM gains a good PA efficiency and thus energy-efficient due to its constant modulus feature.

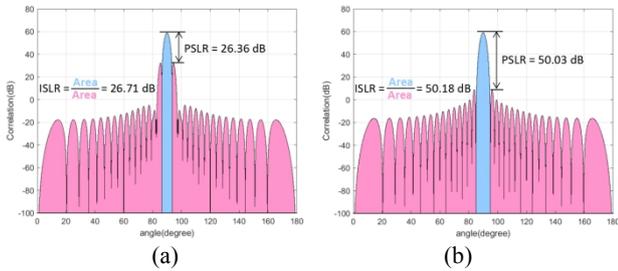

Fig. 4 A PSLR and ISLR comparison of the (a) spatial domain receiving window and (b) spatial domain matched window.

Moreover, since digital OFDM-based ISAC system is employed, the flexibility and efficiency are also very important. In Fig. 3, the conventional numerical integral based NLFM generation [8] and the proposed DNLFM generation are compared. It shows that the accuracy of the conventional NLFM method is still worse than the proposed DNLFM with a 1000-fold generation time. Also, compared with the OFDM-NLFM [9], the proposed DNLFM is generated slightly slower, but the accuracy is much higher, which is improved by around three orders of magnitude.

### B. Spatial-domain PSR and ISR Gain

Fig. 4 shows the PSLR and ISLR before and after adding the spatial-domain matched window. Although there is nearly no SNR gain using the matched window in the spatial domain, it is still very essential to employ the matched window to achieve high PSLR and ISLR in the spatial domain. When using matched windows at both the transmitting and receiving sides, both PSLR, and ISLR can be more than 20 dB higher than those of schemes using only receiving windows, which greatly improves the spatial domain beam pattern and detection performance.

### VI. CONCLUSIONS

In sensing detection, windowing is necessary to suppress the sidelobe of the strong targets to make the weak target distinguishable. This paper aims to increase energy efficiency via adjusting the transmit signal. Unlike conventional communication reference signals, the time-frequency matched windows are added which gains almost 3 dB SNR gain. Also, a novel DNLFM waveform is proposed for the OFDM-based ISAC system, which can be quickly and accurately generated via fast convergence of Newton iterations and geometrical equivalence. The spatial-domain matched windows are also proposed to achieve high PSLR and ISLR. The simulations verify the technical effect of the proposed methods and show a valid performance gain.